\documentclass[showpacs,preprintnumbers,amsmath,amssymb, nofootinbib, prd]{revtex4}
\usepackage{amsmath,amssymb,graphicx}
\usepackage{physics}
\usepackage{epstopdf}
\usepackage {amssymb}
\newcommand{\nc}{\newcommand}
\nc{\ba}{\begin{eqnarray}}
\nc{\ea}{\end{eqnarray}}
\newcommand\be{\begin{equation}}
\newcommand\ee{\end{equation}}

\nc{\x}{{\bf{x}}}

\begin{document}

%%%%%%%%%%%%%%%%%%%%%%%%%%%%%%%%%%%%%%%%%%%%%%%%%%%%%%%%%%%%%%%%%%%%%%%
\title{The effect of Quintessence on the low and high temperature behavior of \\ 
 entanglement entropy for boundary field theory dual to AdS black holes}

\author{A. Farokhtabar}
\email{a.farrokhtabar@stu.umz.ac.ir}
\author{J. Sadeghi}
\email{pouriya@ipm.ir}
\affiliation{Sciences Faculty, Department of Physics,
University of Mazandaran, Babolsar, Iran
\\
P.O.Box 47416-95447.}

\date{\today}

\begin{abstract}
\vspace{0.3cm}

The effect of dark energy on the low and high temperature behavior of strip 
region of the boundary field theory dual to AdS black holes was studied.
In this framework, we investigate this behavior for different types of equation state
 and different regims of normalization factor ($a$). 
We will see that the changes of entanglement entropy
is growing up with incrasing $a$ and $|\omega|$.

\vspace{0.3cm}

\end{abstract}

\maketitle

%%%%%%%%%%%%%%%%%%%%%%%%%%%%%%%%%%%%%%%%%%%%%%%%%%
\section{Introduction}

In recent decades, quantum entanglenent has played an important role in various areas of physics such as quantum 
field theory, condensed matter systems, statistical mechanics and quantum gravity. The entanglement entropy provides
us with a convenient way to measure quantum correlations in a biparite system. We assume that the system is in pure state $\ket{\Psi}$ with the density matrix $\rho=\ket{\Psi}\bra{\Psi}$, therefore von-Neuman entropy of this system is defined as $S=-\mathrm{tr}\rho \log \rho=0$. This quentity does not give us any useful information. Thus the total system is divided two subsystems  $A$ that is bieng studied and $B$ that is the complement of $A$. The observer only can access to information from subsystem $A$ and can not recieve any signal from $B$. This situatioin is analogous to the case that inside of a black hole ($B$) is not accessible for an observer in outside of the horizon ($A$). Mathematically this can be realised by expressing the full Hilbert space $\mathcal{H}$ as a tensor product of Hilbert
spaces $A$ and $B$, that is, $\mathcal{H}=\mathcal{H}_{A}\otimes \mathcal{H}_{B}$. The entanglement entropy of the system $A$ is defined as 
\begin{equation}
S_{A}=-\mathrm{tr}_{A}\rho_{A} \log \rho_{A}
\end{equation}
where $\rho_{A}$ is the reduced density matrix of subsystem $A$ obtained via taking a partial trace $\rho$ over the subststem $B$, $\rho_{A}=\mathrm{tr}_{B}\rho$. The entanglement entropy  of $B$ can be obtained by the same method too.
Entanglement entropy has important properties such as $(i) S_{A}=S_{B}$ and $(ii) S_{A}+S_{B}\geq S_{A \cup B}+S_{A \cap B}$ called subadditivity condition \cite{Araki:1970ar}.
\\
Direct calculation of von-Neuman entropy in quantum field theory is very complicated. Therefore, a method has been propsed based on replica trick \cite{Holzhey:1994we,Callan:1994py,Calabrese:2004eu,Calabrese:2009qy}. In their approach the entanglement entropy is given by
\begin{equation}
S_{n}= \lim_{n \to 1}
\dfrac{1}{1-n} \mathrm{tr} \rho_{A}^{n}
=-\dfrac{\partial}{\partial n}\mathrm{tr} \rho_{A}^{n}|_{n=1}.
\end{equation}
Similar to quantum mechanics, the density matrix is defined in terms of euclidean path integral on an
$n$-sheeted Riemann surface. This method is used for calculating the entanglement entropy of $1+1$ 
CFT in critical \cite{Holzhey:1994we,Vidal:2002rm} and non critical  \cite{Calabrese:2004eu,Calabrese:2009qy} phenamena. Because of the large size of hilbert spaces, replica trick method for calculating entanglement entropy  in higher dimensional CFTs encounters problems. Therefore, in spit of noticable successes for $1+1$ CFT, the direct calculation of entanglement entropy in higher dimentional CFTs is restricted to quasi free fermions and bosons \cite{Cramer:2005mx,Cramer:2005ce,Cramer:2006fu,Plenio:2004he,Gioev:2006hl,Wolf:2006zzb}.
\\
A very beneficial and applicable approach, at least for quantum systems that have holographic description, is using of holographic entanglement entropy that proposed and developed by the authors in \cite{Ryu:2006bv,Hubeny:2007xt}. This method is derived from AdS/CFT duality that relates gravity on an asymptotically local ($d+1$) dimensional AdS
spacetime to ($d$) dimensional strongly coupled boundary quantum field theory with a UV fixed point \cite{Maldacena:1997re}. In the next section, this method is described in ($3+1$ dimensions in details.
\\
 Recently astronomical observations show that the expansion of the univers is accelerated \cite{Perlmutter:1998np,Riess:1998cb},
 that can be explained by the assumption that our universe is filled with a special state of matter with negative pressure. This can be interpretted
 by a cosmological constant, but the measured vaccum energy density differs largely from zero point energy predicted by quantum field theory.
 This disagreement is called cosmological constant problem \cite{Nobbenhuis:2004wn}. An alternative way may be constructing black hole solutions with quintessence by using dynamical scalar fields \cite{Ratra:1988fu,Peebles:1988th,Wetterich:1988rn,Caldwell:1997ii,Zlatev:1998tr}.
 In this framework, a state equation is obtained as a relation between the pressure and energy density. A static spherically-symmetric exact black hole solutions of Einstein equations with the quintessential matter was proposed by Kiselev in \cite{Kiselev:2002dx}. Thermodynamics of the neutral and charged black holes with quintessence was studied in \cite{Ghaderi:2016dpi,Ghaderi:2016ttd,Thomas:2012zzc}. The holographic entanglement entropy was used as a nonlocal observable for probing phase structure in quintessence Reissner-Nordstr\"{o}m-AdS black hole in \cite{Zeng:2015wtt}. From the dual CFT perspective, the effect of quintessence on the formation of superconductor \cite{Chen:2012mva} and on the non-equilibrium thermalization \cite{Zeng:2014xza} were studied. althogh at present, the dual field interpretation of the quintessence is not completely known. In this work, we intend to study the behavior of the entanglement entropy of strongly coupled boundary field theory dual to AdS black holes surrounded by quintessence at low and high temerature.
 \\ 
The rest of this paper is organized as follows.
In section \ref{sec:review} we present a review of the holographic entanglement entropy method for a general metric.
 In the two next sections we calculate the entanglement entropy of AdS black hole surrounded by quintessence at low and high temperature 
 for small and large quintessence charge regims. Finally in the last section, the results summarized as conclusions
 
 \section{\label{sec:review} Review of the holographic entanglement entropy}
 \begin{figure}[bt]
\includegraphics[scale=0.5]{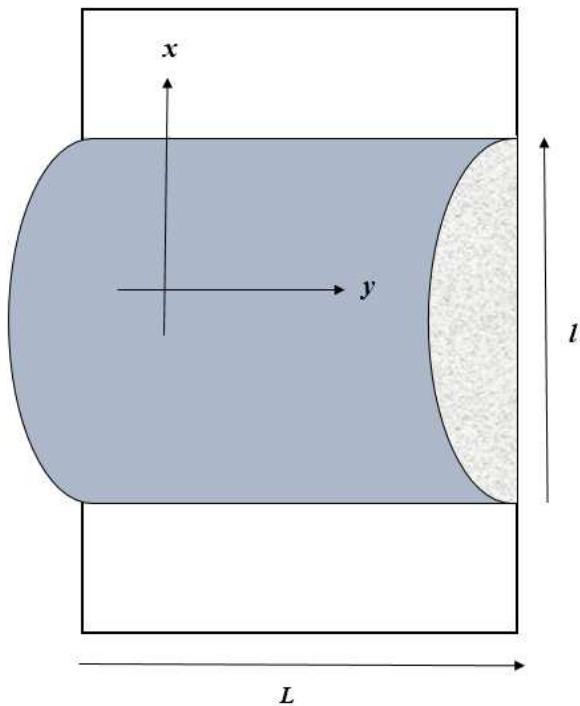}
\caption{The schematic of minimal surface in $AdS_{3+1}$}
\end{figure}
In the first step we review the holographic entanglement entropy method proposed by Ryu and Takayanagi \cite{Ryu:2006bv} for boundary CFT dual to a static spherically-symmetric $AdS_{3+1}$  black hole. As mentioned in previous section, a part of the boundary field theory $A$ is isolated from the rest of system that entangled with $A$ as shown in Fig 1. This subsystem is entangled withthe rest of subsystem.
 According to this proposal, the entanglement entropy is given by
\begin{equation}
\label{eq:holoentropy}
S_{A}=\dfrac{\mathcal{A}(\gamma_{A})}{4G_{N}^{(3+1)}},
\end{equation}
where $G_{N}^{(3+1)}$ is the Newton's constant in $3+1$ dimensional bulk spactime and  $\gamma_{A}$
is the ($2$) dimensional minimal area surface in the bulk whose boundary on the bulk is the boundary of 
conformal field theory on region $A$, $\partial \gamma_{A}=\partial A$. This formula is motivated from
Bekenstein-Hawking  entropy due to similarity between entangled quantum systems and black holes. Although,
the entanglement entropy is different from Bekenstein-Hawking entropy because $\mathrm{Area}(\gamma_{A})$
is not necessarily the area of event horizon. This represents the surface of regin $A$ lives in the conformal boundary 
of AdS spacetime. The entanglement entropy of the AdS Schwarzschild black hole \cite{Cadoni:2009tk,Cadoni:2010sg,Fischler:2012ca,Hubeny:2012ry} and Reissner-Nordstr\"{o}m-AdS black hole 
\cite{Chaturvedi:2016kbk,Kundu:2016dyk} at finite temperature were studied by this method. 
\\
The metric for a black hole in poincar\'{e} coordinte can be written as
\begin{equation}
\label{eq:metric}
ds_{3+1}^{2}=-r^{2}f(r)dt^{2}+\dfrac{1}{r^{2}f(r)}dr^{2}+r^{2}\big( dx^{2}+ dy^{2}\big),
\end{equation}
where we assume the AdS length is set to one. The boundary field theory $A$ is geometrically an infinite strip specified by 
\begin{equation}
\label{eq:boundary}
x\equiv \in[-\frac{l}{2},\frac{l}{2}], \quad y\in[-\frac{L}{2},\frac{L}{2}] 
\end{equation}
with $L \rightarrow \infty$. The minimal area of the surface enclosed by the boudary of $A$ is given by
\begin{equation}
\label{eq:area-formula}
\mathcal{A}=L\int_{-\frac{l}{2}}^{\frac{l}{2}}dr \, r\sqrt{r^{2}x^{\prime 2}+\dfrac{1}{r^{2}f(r)}}.
\end{equation}
If we consider the area as an action, we can see that the lagrangian do not depend to $x$. 
Therefore the equation of motion leads to a constant of motion
\begin{equation}
\label{eq:const-motion}
\dfrac{r^{2}}{\sqrt{1+\dfrac{1}{r^{4}x^{\prime 2}f(r)}}}=const.=c,
\end{equation}
thus, we have
\begin{equation}
\label{eq:dx}
\frac{dx}{dr}=\dfrac{c}{r^{4}\sqrt{\Big(1-\dfrac{c^{2}}{r^{4}}\Big)f(r)}}.
\end{equation}
The costant $c$ may be found by taking this fact that at turning point $r_{c}$, $x^{\prime}$ diverges, therefore, 
we obtain $c^{2}=r_{c}^{4}$. By taking $x(\infty)=\pm \frac{l}{2}$ one can easily derive the width of strip or
extremal length as
\begin{eqnarray}
\label{eq:stripwith}
\frac{l}{2}&=&\int_{r_{c}}^{\infty}dr \, \dfrac{r_{c}^{2}}{r^{4}\sqrt{1-\big(\dfrac{r_{c}}{r}\big)^{4}}}f(r)^{-\frac{1}{2}}
\nonumber \\
&=&\int_{0}^{1}du \, \dfrac{u^{2}}{r_{c}\sqrt{1-u^{4}}}f(u)^{-\frac{1}{2}}
\end{eqnarray}
where we used the variable change $u=\frac{r_{c}}{r}$. The area of the minimal surface can be expressed as
\begin{eqnarray}
\label{eq:minimalarea}
\mathcal{A}&=&2L\int_{r_{c}}^{\infty}dr \, \dfrac{r^{2}}{\sqrt{1-\big(\dfrac{r_{c}}{r}\big)^{4}}}f(r)^{-\frac{1}{2}}
\nonumber \\
&=& 2L\int_{0}^{1}du \, \dfrac{r_{c}}{u^{2}\sqrt{1-u^{4}}}f(u)^{-\frac{1}{2}}.
\end{eqnarray}
When $r \rightarrow \infty$, the intergrand becomes infinte, therefore, the entanglement entropy has a divergence.
Because of too many degrees of freedom, always the entanglement entropy is infinite. For regularization this 
the entanglement entropy can be divided two parts, a divergene part and a fininte part. Divergence part is temperature 
independent and can be computed easily by introducing an infrared cut off $r_{b}$ (dual to UV cut off $a=\frac{1}{r_{b}}$ 
in boundary field). On the other hand, we use the fininite part of entanglement entropy to explore the high and low temparature of 
entanglement entropy for the boundary field dual to AdS black  hole
\begin{equation}
\label{eq:finiteEE}
S_{A}^{\rm{finite}}=S_{A}-S_{A}^{\rm{div}}=\dfrac{\mathcal{A}^{\rm{finite}}}{4G_{N}^{3+1}}.
\end{equation}
The integrals given in equations \eqref{eq:stripwith} and \eqref{eq:minimalarea} had not any analytical solutions yet.
Therefore, a few approximation methods have been proposed for the computation of these integrals, although the most of these methods
are valid for low temperature regime. The authors in \cite{Fischler:2012ca} present a technique based on expansion by which one can calculate
the entanglement entropy of a black hole at low and high temperature. In this paper we apply this technique to obtain an analytical expression
for entanglement entropy of a boundary CFT dual to AdS black hole sorrounded by a quintessence. 
\\
The low temperature limit of boundary CFT corresponds to a black hole with small radius.
It is shown in \cite{Fischler:2012ca} that the leading contribution of entanglement entropy is duo to 
 pure $AdS_{3+1}$ bulk. In this framework the low temperature limit of entannglement entropy of
 boundary CFT dual to AdS black hole can be expressed as
 \begin{equation}
 \label{eq:Entropy-low-ads}
 S_{A} = S_{A}^{AdS}+k(r_{h}l)^{3}.
 \end{equation}
 On the other hand it was shown that at high temperature, that is when $r_{h} \rightarrow r_{c}$,
  the entanglement entropy behaves as the following equation
  \begin{eqnarray}
  \label{eq:Entropy-high-ads}
  S_{A} &=&c_{0}LlT^{2}+Tl(c_{1}+c_{2}\epsilon)+\mathcal{O}[\epsilon^{2}],
  \\
   \label{eq:epsilon-high-ads}
  \epsilon &\approx& \epsilon_{ent} e^{-CTl},
  \end{eqnarray}
  where $c_{i} (i=0,1,2)$ and $\epsilon_{ent}$ and $C$ are constants. In eq. \eqref{eq:Entropy-high-ads}
  the first term scales with area of the subsystem (in higher dimension with volume of the subsystem) and 
  corresponds to thermal entropy of the system. The other terms scales with the length of boundary of subsystem
  (in higher dimension area of the substem) and correspond to the entanglement between the subsystem $A$ and
  the rest of the system.
\section{\label{sec:blakholesol}The entanglement entropy of quintessence planar AdS black holes}
In this section, we firstly present the metric describing Ads planar black holes sorrounded by quintessence.
Then we use Ryu and Takyanagi approach to compute the entanglement entropy of quintessence Ads black hole
dual to a strip region in the boundary field theory. Similar to method used in \cite{Kiselev:2002dx} and considerig 
metric anstaz in eq. \eqref{eq:metric} one can find the function $f(r)$ as \cite{Chen:2008ra} used a technique based
on expansion that 
\begin{equation}
\label{eq:f}
f(r)=1-\frac{2M}{r^{3}}-\dfrac{a}{r^{3(\omega +1)}}
\end{equation}
where $M$ is the mass parameter of the black hole, $\rho$ is the quintessence energy density, $\omega$ is 
state equation parameter and $a$ is the normalization factor that we call it quintessence charge. 
It relates to density of quintessence as follows
\begin{equation}
\label{eq:quintdensity}
\rho=-\dfrac{3\omega \, a}{2r^{3(\omega+1)}}
\end{equation} 
where for quintessence matter $-1<\omega<0$ and $\omega<-1$ is corresponded to phantum dark energy. 
For $\omega=-1$, the qunitessence affects the AdS radius that is due to cosmological constant while for 
$\omega=-\frac{1}{3}$, dark energy affects the curvature $k$ of the spacetime.
\\
For deriving the relation between $M$ and $a$, it is sufficient to put $f(r_{h})=0$ where $r_{h}$
is the horizon radius of the black hole, therefore, we have
\begin{equation}
\label{eq:mass}
M=\frac{1}{2}\big(r_{h}^{3}-\frac{a}{r_{h}^{3\omega}}\big).
\end{equation}
Here, the positive mass of black hole leads us to have the following condition
\begin{equation}
\label{eq:cond}
\frac{a}{r_{h}^{3(1+\omega)}}<1.
\end{equation}
$f(r)$ can be reformulated in term of the horizon radius as
\begin{equation}
\label{eq:frh}
f(r)=1-(\frac{r_{h}}{r})^{3}+\frac{a}{r^{3}}\big(\frac{1}{r_{h}^{3\omega}}-\frac{1}{r^{3\omega}}\big)
\end{equation}
\\
The Hawking temperature of the quintessence AdS blach holes can be written as
\begin{equation}
\label{eq:Hawking-temp}
T=\frac{r^{2}f^{\prime}(r)}{4\pi}\Big\vert_{r=r_{h}}=\frac{3 \, r_{h}}{4\pi}\big(1+\frac{a\, \omega}{r_{h}^{3(\omega +1)}}\big).
\end{equation}
The allowed range of black hole temperature can be obtained by considering condition \eqref{eq:cond} as
\begin{equation}
\label{eq:Temp-cond}
\dfrac{3r_{h}}{4\pi}(1+\omega)<T<\dfrac{3r_{h}}{4\pi}
\end{equation}
By substituting $f(r)$ from eq. \eqref{eq:frh} in equations \eqref{eq:stripwith} and\eqref{eq:area-formula}, 
$l$ and $\mathcal{A}$ are ghiven by
\begin{eqnarray}
\label{eq:l}
l &=&\frac{2}{r_{c}}\int_{0}^{1}du \, \dfrac{u^{2}}{\sqrt{1-u^{4}}}
\Big( 1-(\frac{r_{h}}{r_{c}})^{3}u^{3}+\frac{a}{r_{h}^{3\, \omega} r_{c}^{3}}u^{3}
-\frac{a}{r_{c}^{3(\omega+1)}}u^{3(\omega +1)}
\Big)^{-\frac{1}{2}},
\\
\label{eq:A}
\mathcal{A}&=& 2L\int_{0}^{1}du \, \dfrac{r_{c}}{u^{2}\sqrt{1-u^{4}}}
\Big( 1-(\frac{r_{h}}{r_{c}})^{3}u^{3}+\frac{a}{r_{h}^{3\, \omega} r_{c}^{3}}u^{3}
-\frac{a}{r_{c}^{d(\omega+1)}}u^{3(\omega +1)}
\Big)^{-\frac{1}{2}}.
\end{eqnarray}
For a given $\omega$, the state space of the boundary field theory dual to quintessence AdS black hole
depends on $T$ and $a$. Therefore it is needed to work in a specific ensemble. In order to fix $a$,
we consider the system in canonical ensemble. Thus, in our study for finite temperature entanglement entropy, 
by attention to \eqref{eq:Hawking-temp} one can find that the value of 
$r_{h}$ can be considered as a measure of the black hole 
tempreature. Except in the cases $\omega$ is very close to $-1$, the large quintessence charge $a$
, $r_{h}$ must be large, the larger the radius
of black hole, the higher the temperature. We must notice that when $\omega \rightarrow -1$, a large value of $a$, the large value of $r_{h}$ is not necessarily related to high temperature. For small values of $a$, the radius of black hole can be small or large. This can be related to low or high temperature respectively
\\
In the following subsections we study the behavior of entanglement entropy for the boundary CFT dual to AdS
black hole surrounded by quintessence at different quintessence charge regims.
 \section{\label{Small charge} Black Hole in Quintessence with a small charege}
 In this section we explore the low and high temperature behavior of  entanglement 
 entropy for subsystem $A$ boundary to AdS black hole sourrunded by quintessence.
 We study these bahaviors at different region of $\omega$.
\subsection{\label{susec:non-small a}Small $a$ regime (low temperature)}
At first step we consider a subsystem of boundary field theory dual to AdS quintessence black hole with a small $a$
parameter at a low temperature. From the mass condition in eq. \eqref{eq:cond}, we can find that
when the temperature and the charge both are small, the horizon radius is small. In this condition $r_{h}\ll r_{c}$.
 We can find that when $\omega$ is not very close to $-1$, the small temperature and quintessence charge leads to 
a relation as $\frac{a}{r_{h}^{3(\omega +1)}} \sim 1$ Therefore, in this case the lapse function $f(u)$ has a form as
\begin{equation}
\label{eq:f-n-ext}
f(u)=1-\big(\frac{r_{h}}{r_{c}}\big)^{3}u^{3}
+\frac{a}{r_{h}^{3(\omega +1)}}\Big[
\big(\frac{r_{h}}{r_{c}}\big)^{3}u^{3}-
\big(\frac{r_{h}}{r_{c}}\big)^{3(\omega +1)}u^{3(\omega +1)}
\Big]
\end{equation}
Defining a new parameter $\xi =\frac{a}{r_{h}^{3(\omega +1)}}$, we expand $f^{-\frac{1}{2}}(u)$
around $\frac{r_{h}}{r_{c}}=0$. Therefore by keeping the terms up to $\mathcal{O}([\frac{r_{h}}{r_{c}}]^{3}u^{3})$,
the lapse function has the following form
\begin{equation}
\label{eq:f-n-ext1}
f^{-\frac{1}{2}}(u) \approx 1+\frac{1-\xi}{2}
\big(\frac{r_{h}}{r_{c}}\big)^{3}u^{3}
+\sum_{n=1}^{\lfloor\frac{1}{\omega+1}\rfloor}
\dfrac{\Gamma(n+\frac{1}{2})}{\sqrt{\pi}\Gamma(n+1)}\xi^{n}
\big(\frac{r_{h}}{r_{c}}\big)^{3n(\omega+1)}u^{3n(\omega+1)}
\end{equation}
Substituting this approximation form of lapse function in \eqref{eq:stripwith}
the subsystem length can be expressed by the following integral form
\begin{equation}
\label{eq:l-nonextint}
l \approx \frac{2}{r_{c}}\int_{0}^{1}\dfrac{u^{2}du}{\sqrt{1-u^{4}}}\Big(
1+\frac{1-\xi}{2}\big(\frac{r_{h}}{r_{c}}\big)^{3}u^{3}
+\sum_{n=1}^{\lfloor\frac{1}{\omega+1}\rfloor}
\dfrac{\Gamma(n+\frac{1}{2})}{\sqrt{\pi}\Gamma(n+1)}\xi^{n}
\big(\frac{r_{h}}{r_{c}}\big)^{3n(\omega+1)}u^{3n(\omega+1)}
\Big)
\end{equation}
we can solve above integral and derive the relation between $l$ and $r_{c}$ as
\begin{equation}
\label{eq:rc-nonext}
r_{c} \approx \frac{2}{l}\Big[
\dfrac{\sqrt{\pi}\Gamma \big(\frac{3}{4}\big)}{\Gamma \big(\frac{1}{4}\big)}
+\sum_{n=1}^{\lfloor\frac{1}{\omega+1}\rfloor}
\dfrac{\Gamma(n+\frac{1}{2})\Gamma \Big(\dfrac{3n(\omega+1)+3}{4}\Big)}
{4\Gamma(n+1)\Gamma\Big(\dfrac{3n(\omega+1)+5}{4}\Big)}\xi^{n}
\big(\frac{r_{h}}{r_{c}}\big)^{3n(\omega+1)}
+\dfrac{\pi(1-\xi)}{16}
\big(\frac{r_{h}}{r_{c}}\big)^{3}\Big]
\end{equation}
The extremal area surface can be obtained as
\begin{equation}
\label{eq:A-nonextint}
\mathcal{A} \approx 2Lr_{c}\int_{0}^{1}\dfrac{du}{u^{2}\sqrt{1-u^{4}}}\Big(
1+ \sum_{n=1}^{\lfloor\frac{1}{\omega+1}\rfloor}
\dfrac{\Gamma(n+\frac{1}{2})}{\sqrt{\pi}\Gamma(n+1)}\xi^{n}
\big(\frac{r_{h}}{r_{c}}\big)^{3n(\omega+1)}u^{3n(\omega+1)}
+\frac{1-\xi}{2}\big(\frac{r_{h}}{r_{c}}\big)^{3}u^{3}\Big)
\end{equation}
the first term in \eqref{eq:A-nonextint} is correspond to pure AdS and is divergent. Therefore, in order to obtain a finite area 
we introduce an IR cut off $r_{b}$ (dual to UV cut off $c=\frac{1}{r_{b}}$ of boundary theory) 
and subtract a counter term from it.Therefore the first term can be regularized as
\begin{equation}
\label{eq:A1-finite}
\mathcal{A}^{finite}_{1}= 2Lr_{c}\int_{\frac{r_{c}}{r_{b}}}^{1}du \, 
\dfrac{1}{u^{2}\sqrt{1-u^{4}}}-2Lr_{b} 
=\dfrac{\sqrt{\pi}L \, r_{c}}{2}\dfrac{{\Gamma \big( -\dfrac{1}{4}\big)}}
{{\Gamma \big( \dfrac{1}{4}\big)}}
\end{equation}
The terms in the sum are divergent when for a special $n$ in the sum, 
$n<\frac{2}{3(1+\omega)}$. Therefore when in the sum $n<\frac{2}{3(1+\omega)}$ and $3n(\omega+1)\neq1$
The integral for these term can be solved as follows
\begin{eqnarray}
\label{eq:A2ext}
\mathcal{A}^{finite}_{2}&=& 2Lr_{c}
\sum_{n=1}^{\lfloor\frac{1}{\omega+1}\rfloor}
\dfrac{\Gamma(n+\frac{1}{2})}{\sqrt{\pi}\Gamma(n+1)}
\Big[\int_{\frac{r_{c}}{r_{b}}}^{1}du \, 
\dfrac{u^{3n(\omega+1)-2}}{\sqrt{1-u^{4}}} \nonumber
\\
&+&\sum_{\substack{n<\frac{2}{3(1+\omega)} \\
3n(\omega+1)\neq1}}
\frac{1}{3n(\omega +1)-2}\big(\frac{r_{c}}{r_{b}}\big)^{3n(\omega +1)-2}\Big]
\big(\frac{r_{h}}{r_{c}}\big)^{3n(\omega+1)} \nonumber \\
&=& 
\dfrac{L \, r_{c}}{2}\sum_{n=1}^{\lfloor\frac{1}{\omega+1}\rfloor}
\dfrac{\Gamma(n+\frac{1}{2})}{\Gamma(n+1)}
\dfrac{{\Gamma \big( \dfrac{3n(\omega+1)-1}{4}\big)}}
{{\Gamma \Big( \dfrac{3n(\omega+1)+1}{4}\Big)}}
\big(\frac{r_{h}}{r_{c}}\big)^{3n(\omega+1)} 
\end{eqnarray}
Therefore when in the sum $3n(\omega+1) \neq 1$ the extremal area can be expressed as
\begin{equation}
\label{eq:nonext-A}
\mathcal{A}^{finite} = \dfrac{L \, r_{c}}{2}
\Big[\dfrac{\sqrt{\pi}\Gamma \big( -\dfrac{1}{4}\big)}
{\Gamma \big( \dfrac{1}{4}\big)} \nonumber
+\sum_{n=1}^{\lfloor\frac{1}{\omega+1}\rfloor}
\dfrac{\Gamma(n+\frac{1}{2})}{\Gamma(n+1)}
\dfrac{{\Gamma \Big( \dfrac{3n(\omega+1)-1}{4}\Big)}}
{{\Gamma \Big( \dfrac{3n(\omega+1)+1}{4}\Big)}}\xi^{n}
\big(\frac{r_{h}}{r_{c}}\big)^{3n(\omega+1)}
+ \frac{\pi}{2}\big(1-\xi\big)
\big(\frac{r_{h}}{r_{c}}\big)^{3}\Big]
\end{equation}
When $\omega \rightarrow -1$, we have many terms in the lapse function \eqref{eq:f-n-ext}
 and the calculation is very hard and almost imposible.
 Fortunately, in this case, $r_{h}^{3(\omega+1)} \rightarrow 1$.
 Thus $\xi \ll 1$ and we can neglect the higher order terms in the lapse function.
 $f^{-\frac{1}{2}}$ has a form as follows
\begin{equation}
\label{eq:lapse-asmall-Tlow-w1}
f^{-\frac{1}{2}}(u) \approx 1+\frac{1}{2}\xi
\big(\frac{r_{h}}{r_{c}}\big)^{3(\omega+1)}
u^{3(\omega+1)}+\frac{1-\xi}{2}
\big(\frac{r_{h}}{r_{c}}\big)^{3}u^{3}
\end{equation}
Therefore the subsystem length and minimal surface area are expessed as follow
\begin{eqnarray}
\label{eq:Tlow-asmal-w1-l}
r_{c} &\approx & \frac{2}{l}\Big[
\sqrt{\pi}\dfrac{\Gamma (\frac{3}{4})}{\Gamma (\frac{1}{4})}
+\frac{1}{8}\sqrt{\pi}\dfrac{\Gamma \big(\frac{3\omega+6}{4}\big)}
{\Gamma \big(\frac{3\omega+8}{4}\big)}\xi \big(r_{h}l\big)^{3(\omega+1)}
-\frac{1}{2\sqrt{\pi}}(1-\xi)\dfrac{\Gamma (\frac{1}{4})^{3}}
{\Gamma (-\frac{1}{4})^{3}}\big(r_{h}l\big)^{3}\Big]
\\
\label{eq:Tlow-asmal-w1-A}
\mathcal{A}^{finite} &\approx& \frac{L}{l}\Big[-\frac{\pi}{4
}\dfrac{\Gamma (-\frac{1}{4})^{2}}
{\Gamma (\frac{1}{4})^{2}}+\frac{1}{2}\dfrac{\Gamma (\frac{1}{4})^{2}}
{\Gamma (-\frac{1}{4})^{2}}(1-\xi)\big(r_{h}l\big)^{3}
+\frac{1}{2}\pi \dfrac{\Gamma (-\frac{1}{4})}
{\Gamma (\frac{1}{4})}\Big \{
\dfrac{\Gamma \big(\frac{3\omega+6}{4}\big)}
{4\Gamma \big(\frac{3\omega+8}{4}\big)}
-\dfrac{\Gamma \big(\frac{3\omega+2}{4}\big)}
{2\Gamma \big(\frac{3\omega+4}{4}\big)}\Big \}
\xi \big(r_{h}l\big)^{3(\omega+1)}
\Big]
\end{eqnarray}
Here we compute the entanglement entropy for several values of $\omega$.
For $\omega=-\frac{1}{3}$, $r_{c}$ and $\mathcal{A}$ are expressed as
\begin{eqnarray}
\label{eq:rc-nonextw13-low}
r_{c} &=& \frac{2}{l}\dfrac{\sqrt{\pi}\Gamma \big(\frac{3}{4}\big)}{\Gamma \big(\frac{1}{4}\big)}\Big[
1+\frac{\xi}{96 \pi}\dfrac{\Gamma \big(\frac{1}{4}\big)^{4}}{\Gamma\big(\frac{3}{4}\big)^{4}}\big(r_{h}l\big)^{2}
+\frac{1-\xi}{128 \pi}\dfrac{\Gamma \big(\frac{1}{4}\big)^{4}}{\Gamma\big(\frac{3}{4}\big)^{4}}\big(r_{h}l\big)^{3}
+ \mathcal{O} \big(r_{h}^{4}l^{4}\big) \Big]
\\
\label{eq:A-nonextw13-low}
\mathcal{A}^{finite} &=& -\frac{1}{4}\big(\frac{L}{l}\big)\pi \dfrac{\Gamma \big(-\frac{1}{4}\big)^{2}}
{\Gamma \big(\frac{1}{4}\big)^{2}}\Big[
1-\frac{16\xi}{3\pi}\dfrac{\Gamma \big(\frac{1}{4}\big)^{4}}{\Gamma\big(-\frac{1}{4}\big)^{4}}\big(r_{h}l\big)^{2}
-\frac{2(1-\xi)}{\pi}\dfrac{\Gamma \big(\frac{1}{4}\big)^{4}}{\Gamma\big(-\frac{1}{4}\big)^{4}}\big(r_{h}l\big)^{3}
+ \mathcal{O} \big(r_{h}^{4}l^{4}\big) \Big]
\end{eqnarray}
For $\omega=-\frac{1}{2}$, these parameters can be expressed as
\begin{eqnarray}
\label{eq:rc-nonextw12-low}
r_{c}&=&\frac{2}{l}\Big[C_{0}+C_{1}(r_{h}l)^{\frac{3}{2}}
+C_{2}(r_{h}l)^{3}+\mathcal{O}\big([r_{h}l)^{\frac{9}{2}}\Big]
\\
\label{eq:A-nonextw12-low}
\mathcal{A}^{finite}&=&\frac{L}{l}\Big[K_{0}+K_{1}(r_{h}l)^{\frac{3}{2}}
+K_{2}(r_{h}l)^{3}+\mathcal{O}\big([r_{h}l)^{\frac{9}{2}}\Big]
\end{eqnarray}
where the cofficients $C_{i}$ and $K_{i}$ ($i=0,1,2$) are given by the expressions below
\begin{eqnarray}
\label{eq:nonlowc0}
C_{0}&=&\sqrt{\pi}\dfrac{\Gamma (\frac{3}{4})}{\Gamma (\frac{1}{4})}\\
\label{eq:nonlowc1}
C_{1} &=& -\dfrac{2^{\frac{9}{4}}\pi^{\frac{5}{4}}\xi}{5}
\dfrac{\Gamma (\frac{1}{8})}{\Gamma (-\frac{1}{4})^{3}\Gamma (\frac{5}{8})}\\
C_{2} &=& -\dfrac{3\xi^{2}}{100\pi}
\dfrac{\Gamma (\frac{1}{8})^{2}\Gamma (\frac{1}{4})^{4}}
{\Gamma (-\frac{1}{4})^{4}\Gamma (\frac{5}{8})^{2}}
-\dfrac{3\xi^{2}}{8\sqrt{\pi}}
\dfrac{\Gamma (\frac{1}{4})^{3}}
{\Gamma (-\frac{1}{4})^{3}}
-\dfrac{1-\xi}{2\sqrt{\pi}}
\dfrac{\Gamma (\frac{1}{4})^{3}}
{\Gamma (-\frac{1}{4})^{3}}
\\
K_{0}&=& -\frac{\pi}{4}\dfrac{\Gamma (-\frac{1}{4})^{2}}
{\Gamma (\frac{1}{4})^{2}}
\\
K_{1}&=& \dfrac{2^{\frac{1}{4}}\pi^{-\frac{1}{4}}\xi}{10}\dfrac{\Gamma (\frac{1}{8})
\Gamma (\frac{1}{4})}{\Gamma (\frac{5}{8})}
\\
K_{2}&=& \frac{1}{2}\dfrac{\Gamma (\frac{1}{4})^{2}}
{\Gamma (-\frac{1}{4})^{2}}(1-\xi)
+\Big(\frac{1}{50\sqrt{\pi}}\dfrac{\Gamma (\frac{1}{8})^{2}\Gamma (-\frac{1}{4})^{3}}
{\Gamma (\frac{1}{4})^{3}\Gamma (\frac{5}{8})^{2}}
+\frac{3}{4}\dfrac{\Gamma (\frac{1}{4})^{2}}
{\Gamma (-\frac{1}{4})^{2}}\Big)\xi^{2}
\end{eqnarray}
For $\omega =-\frac{2}{3}$ the extremal length can be expressed as
\begin{equation}
\label{eq:w23-low}
r_{c}=\frac{2}{l}\Big[c_{0}+c_{1}(r_{h}l)+c_{2}(r_{h}l)^{2}
+c_{3}(r_{h}l)^{3}+\mathcal{O}(r_{h}l)^{4}\Big]
\end{equation}
where the cofficients $c_{i}$, $(i=0,1,2,3)$ can be obtained as
\begin{eqnarray}
\label{eq:c0-w13}
c_{0}&=&\sqrt{\pi}\dfrac{\Gamma (\frac{3}{4})}{\Gamma (\frac{1}{4})}\\
\label{eq:c1-w12}
c_{1}&=&\frac{1}{2\sqrt{\pi}}\dfrac{\Gamma (\frac{1}{4})}{\Gamma (-\frac{1}{4})}\xi\\
\label{eq:c1-w23}
c_{2}&=&(\frac{2}{\pi\sqrt{\pi}}-\frac{1}{\sqrt{\pi}})\xi^{2}\dfrac{\Gamma (\frac{1}{4})^{3}}
{\Gamma (-\frac{1}{4})^{3}}\\
\label{eq:c3-w23}
c_{3}&=&\Big[(\frac{4}{\pi^{\frac{5}{2}}}-\frac{3}{\pi^{\frac{3}{2}}})
\dfrac{\Gamma (\frac{1}{4})^{5}}{\Gamma (-\frac{1}{4})^{5}}
+\frac{5}{8}\dfrac{\Gamma (\frac{1}{4})^{2}}{\Gamma (-\frac{1}{4})^{2}}\Big]\xi^{3}
-\frac{1}{2\sqrt{\pi}}(1-\xi)\dfrac{\Gamma (\frac{1}{4})^{3}}
{\Gamma (-\frac{1}{4})^{3}}
\end{eqnarray}
Therefore the extremal area can be expressed as
\begin{equation}
\label{eq:lowA}
\mathcal{A}^{finite}=\frac{L}{l}\Big[
k_{0}+k_{1}(r_{h}l)+k_{2}(r_{h}l)^{2}+
k_{3}(r_{h}l)^{3}+\mathcal{O}(r_{h}l)^{4}\Big]
\end{equation}
The cofficients are given by the following expressions
\begin{eqnarray}
\label{eq:cofk0}
k_{0}&=&-\frac{\pi}{4}\dfrac{\Gamma (-\frac{1}{4})^{2}}
{\Gamma (\frac{1}{4})^{2}}\\
\label{eq:cofk1}
k_{1}&=& \xi(\log 2-\frac{1}{2})\\
\label{eq:cofk2}
k_{2}&=&\frac{1}{\pi}\dfrac{\Gamma (\frac{1}{4})^{2}}
{\Gamma (-\frac{1}{4})^{2}}\xi^{2}\\
\label{eq:cofk3}
k_{3}&=&\frac{1}{2}\dfrac{\Gamma (\frac{1}{4})^{2}}
{\Gamma (-\frac{1}{4})^{2}}(1-\xi)
+(\frac{2}{\pi}-\frac{4}{\pi^{2}})\xi^{3}\dfrac{\Gamma (\frac{1}{4})^{4}}
{\Gamma (-\frac{1}{4})^{4}}
\end{eqnarray}
For $\omega =-1$, $f^{-\frac{1}{2}}(u)$ has the following form
\begin{equation}
\label{eq:f-w1}
f^{-\frac{1}{2}}(u)=\frac{1}{\sqrt{1-a}}\Big(
1-\big(\frac{r_{h}}{r_{c}}\big)^{3}u^{3}
\Big)^{-\frac{1}{2}}
\end{equation}
The form of the lapse function is the same as that of schwarzchild black holes in \cite{Fischler:2012ca}
with the exception that here there is a $\frac{1}{\sqrt{1-a}}$ factor. For low temperature
$r_{c}$ and $\mathcal{A}$ can be expressed as follow
\begin{eqnarray}
\label{eq:w1-l}
r_{c} &=& \frac{2}{l\sqrt{1-a}}\Big[
\sqrt{\pi}\dfrac{\Gamma (\frac{3}{4})}{\Gamma (\frac{1}{4})}
-\frac{1}{2\sqrt{\pi}}\dfrac{\Gamma (\frac{1}{4})^{3}}
{\Gamma (-\frac{1}{4})^{3}}\big(r_{h}l\big)^{3}
+\mathcal{O}[r_{h}^{6}l^{6}]\Big]
\\
\label{eq:w1-A}
\mathcal{A}^{finite}&=&\frac{L}{l\sqrt{1-a}}\Big[
-\frac{\pi}{4}\dfrac{\Gamma (-\frac{1}{4})^{2}}
{\Gamma (\frac{1}{4})^{2}}
+\frac{1}{2}\dfrac{\Gamma (\frac{1}{4})^{2}}
{\Gamma (-\frac{1}{4})^{2}}(r_{h}l)^{3}
+\mathcal{O}[r_{h}^{6}l^{6}]\Big]
\end{eqnarray}
For small quintessence charge regime, we  have
$\frac{1}{\sqrt{1-a}}\approx 1+\frac{1}{2}a$ and in this limit the equations
\eqref{eq:w1-l} and \eqref{eq:w1-A} can be rewritten as
\begin{eqnarray}
\label{eq::w1-l2}
r_{c} &\approx& \frac{2(1+\frac{1}{2}a)}{l}\Big[
\sqrt{\pi}\dfrac{\Gamma (\frac{3}{4})}{\Gamma (\frac{1}{4})}
-\frac{1}{2\sqrt{\pi}}\dfrac{\Gamma (\frac{1}{4})^{3}}
{\Gamma (-\frac{1}{4})^{3}}\big(r_{h}l\big)^{3}
+\mathcal{O}[r_{h}^{6}l^{6}]\Big]
\\
\label{eq:w1-A2}
\mathcal{A}^{finite}&\approx&\frac{L(1+\frac{1}{2}a)}{l}\Big[
-\frac{\pi}{4}\dfrac{\Gamma (-\frac{1}{4})^{2}}
{\Gamma (\frac{1}{4})^{2}}
+\frac{1}{2}\dfrac{\Gamma (\frac{1}{4})^{2}}
{\Gamma (-\frac{1}{4})^{2}}(r_{h}l)^{3}
+\mathcal{O}[r_{h}^{6}l^{6}]\Big]
\end{eqnarray}
This result is consistent with equation \eqref{eq:Tlow-asmal-w1-A} when $\omega=-1$.
\\
We can see that in all cases the first term in entanglenglent entropy is a constant. This term
is the entanglement entropy of subsystem $A$ when the bulk is pure AdS. Also we can find that
when the quintessence charge is small, the leading contribution to the entanglement entropy of 
the subsystem $A$ dual to AdS black hole sorrounded by quintessence arises from the pure spacetime.
Besides, the effect of quintessence on entanglement entropy is very small even for $\omega=-1$ the
dark energy does not affect noticably the value of entanglement entropy of the subsystem.
\subsection{High temperature regime}
In this subsection we investigate the high temperature behavior of the 
entanglement entropy of the black hole when the parameter $a$ is small.
For high temperature, the horizon raduis is large. Therefore for small quintessence
charge we have $\frac{a}{r_{h}^{3(\omega+1)}} \ll 1$. We define a new parameter as
$\delta=\frac{a}{r_{h}^{3(\omega+1)}}$ and expand $f^{-\frac{1}{2}}(u)$ by Taylor 
series around $\delta=0$ as
\begin{equation}
\label{eq:fhighasmall}
f^{-\frac{1}{2}}(u) \approx \dfrac{1}{\sqrt{1-(\frac{r_{h}}{r_{c}})^{3}u^{3}}}
-\frac{\delta}{2}\frac{r_{h}^{3}}{r_{c}^{3}}u^{3}
\dfrac{\big(1-(\frac{r_{h}}{r_{c}})^{3\omega}u^{3\omega}\big)}
{(1-(\frac{r_{h}}{r_{c}})^{3}u^{3})^{\frac{3}{2}}}
\end{equation}
Substiting the above approximated forms of lapse function in the integral expressions
\eqref{eq:l} and \eqref{eq:A}, we obtain the subsystem length and extremal surface as
\begin{eqnarray}
\label{eq:l-high-asmall}
l &=&\frac{2}{r_{c}}\int_{0}^{1}du \, \dfrac{u^{2}}{\sqrt{1-u^{4}}}
\Big( \dfrac{1}{\sqrt{1-[\frac{r_{h}}{r_{c}}]^{3}u^{3}}}
-\frac{\delta}{2}\frac{r_{h}^{3}}{r_{c}^{3}}u^{3}
\dfrac{\big(1-[\frac{r_{h}}{r_{c}}]^{3\omega}u^{3\omega}\big)}
{(1-[\frac{r_{h}}{r_{c}}]^{3}u^{3})^{\frac{3}{2}}}
\Big),
\\
\label{eq:A-high-asmall}
\mathcal{A}&=& 2L\int_{0}^{1}du \, \dfrac{r_{c}}{u^{2}\sqrt{1-u^{4}}}
\Big( \dfrac{1}{\sqrt{1-[\frac{r_{h}}{r_{c}}]^{3}u^{3}}}
-\frac{\delta}{2}\frac{r_{h}^{3}}{r_{c}^{3}}u^{3}
\dfrac{\big(1-[\frac{r_{h}}{r_{c}}]^{3\omega}u^{3\omega}\big)}
{(1-[\frac{r_{h}}{r_{c}}]^{3}u^{3})^{\frac{3}{2}}}
\Big)
\end{eqnarray}
As showed in \cite{Hubeny:2012ry}, the extremal surface can never penetrate the horizon.
This implies that $r_{c}$ is always larger than $r_{h}$. At High temperature $r_{h}$ is very large
, therefore, it approaches extremal surface, $r_{h} \sim r_{c}$. Assuming $r_{c}=r_{h}(1+\epsilon)$
and  proceeding similar to that used in \cite{Chaturvedi:2016kbk}, the form of boundary subsystem 
length can be obtained as
\begin{equation}
\label{eq:high-asmall-l}
l \, r_{h} = a_{1}+\delta a_{2}-\frac{1}{\sqrt{3}}\log [3\epsilon]+\mathcal{O}[\epsilon]\\
%\label{eq:high-asmall-A}
%\mathcal{A} &=& L \, r_{h}^{2}(1+2\epsilon)+L\, r_{h} \Big[
%b_{1}+\delta b_{2}+b_{3} \epsilon +\delta (b_{4}\epsilon
%+b_{5}\epsilon \log [\epsilon] +\mathcal{O}[\epsilon^{2}])
%\Big]
\end{equation}
where the cofficients $a_{1}$ and $a_{2}$ are expressed as the following forms
\begin{eqnarray}
\label{eq:coffa1}
a_{1} &=& \frac{1}{2}\sqrt{\pi}\frac{\Gamma (\frac{3}{4})}{\Gamma (\frac{5}{4})}
+\sum_{n=1}^{\infty}\Big(\dfrac{\Gamma (n+\frac{1}{2})
\Gamma (\frac{3n+3}{4})}{2\Gamma (n+1)
\Gamma (\frac{3n+5}{4})}-\frac{1}{\sqrt{3}n}\Big)
\\
\label{eq:coffa2}
a_{2} &=& \frac{1}{2}\sum_{n=0}^{\infty}\Big(\dfrac{\Gamma (n+\frac{3}{2})
\Gamma (\frac{3n+3\omega+6}{4})}{\Gamma (n+1)
\Gamma (\frac{3n+3\omega+8}{4})}
-\dfrac{\Gamma (n+\frac{3}{2})
\Gamma (\frac{3n+6}{4})}{\Gamma (n+1)
\Gamma (\frac{3n+8}{4})}-\frac{\omega}{\sqrt{3}}\Big)
\end{eqnarray}
We can obtain $\epsilon$ correction as
\begin{equation}
\label{eq:e-correction}
\epsilon \approx \epsilon_{ent} e^{\frac{-4\pi}{\sqrt{3}}Tl(1-\omega \delta+\frac{a_{2}}{l}\delta)}
\end{equation}
where $\epsilon_{ent}$ is a constant has the following form
\begin{equation}
\label{eq:epsilon-ent}
\epsilon_{ent} \approx \frac{1}{3}\exp (a_{1})
\end{equation}
The $\epsilon$ corrections decreases exponentialy with the temparture 
just as they do in the case Schwarzschild black hole in vaccum. The only difference is
introduction of two small $\delta$ terms duo to the presence of quintessence. This terms cause 
$\epsilon$ to decay in higher temperature rather than Schwarzschild black hole.  
\\
The extremal surface area can be written as
\begin{equation}
\label{eq:high-asmall-A}
\mathcal{A}^{finite} = L \, l r_{h}^{2}(1+2\epsilon)+L\, r_{h} \Big[
b_{1}+\delta b_{2}+b_{3} \epsilon +b_{4} \epsilon \log \epsilon
+\delta (b_{5}\epsilon+b_{6}\epsilon \log [\epsilon] +\mathcal{O}[\epsilon^{2}])
\Big]
\end{equation}
$b_{i} (i=1,2,3,4,5)$ can be obtained as
\begin{eqnarray}
\label{eq:b1}
b_{1}&=&\sqrt{\pi}\frac{\Gamma (-\frac{1}{4})}{\Gamma (\frac{5}{4})}
+\frac{\pi^{2}}{9\sqrt{3}}+\sum_{n=1}^{\infty}\Big(\frac{1}{3n-1}\dfrac{\Gamma (n+\frac{1}{2})
\Gamma (\frac{3n+3}{4})}{\Gamma (n+1)
\Gamma (\frac{3n+5}{4})}-\frac{2}{3\sqrt{3}n^{2}}\Big)
\\
\label{eq:b2}
b_{2} &=& \frac{\sqrt{\pi}}{2+3\omega}\dfrac{\Gamma (\frac{6+3\omega}{4})}
{\Gamma (\frac{8+3\omega}{4})}-\frac{\pi}{8}
+\sum_{n=1}^{\infty}\Big(\frac{1}{3n+2+3\omega}\dfrac{\Gamma (n+\frac{3}{2})
\Gamma (\frac{3n+3\omega+6}{4})}{\Gamma (n+1)
\Gamma (\frac{3n+3\omega+8}{4})}\Big)\nonumber \\
&-&\sum_{n=1}^{\infty}\Big(\frac{1}{3n+2}\dfrac{\Gamma (n+\frac{3}{2})
\Gamma (\frac{3n+6}{4})}{\Gamma (n+1)
\Gamma (\frac{3n+8}{4})}\Big)
\\
\label{eq:b3}
b_{3} &=& \sqrt{\pi}\frac{\Gamma (-\frac{1}{4})}{\Gamma (\frac{1}{4})}
+\frac{\pi^{2}}{9\sqrt{3}}+\frac{2 \log 3}{\sqrt{3}}-\sum_{n=1}^{\infty}
\Big(\dfrac{\Gamma (n+\frac{1}{2})
\Gamma (\frac{3n+3}{4})}{\Gamma (n+1)
\Gamma (\frac{3n+5}{4})}-\frac{6n-2}{3\sqrt{3}n^{2}}\Big)
\\
\label{eq:b4}
b_{4} &=& \frac{2}{\sqrt{3}}\log 3 \\
\label{eq:b5}
b_{5} &=& \frac{\pi}{4}-\dfrac{\sqrt{\pi}\Gamma (\frac{6+3\omega}{4})}
{2\Gamma (\frac{8+3\omega}{4})}\\
\label{eq:b6}
b_{6} &=& \frac{2\omega}{3\sqrt{3}}
\end{eqnarray}
where for computing these cofficients we applied the same method used in \cite{Chaturvedi:2016kbk}.
Therefore the finite part of the entanglement entropy of boundary field theory dual to quintessence black 
hole may be written down as follows
\begin{equation}
\label{eq:S-high-asmall}
S^{fimite}=LlS_{BH}(1+2\epsilon)+\frac{Lr_{h}}{4G}(b_{1}
+\delta b_{2})+\frac{Lr_{h}\epsilon}{4G}\Big[b_{3}+b_{4}\log \epsilon
+\delta \big(b_{5}+b_{6} \log [\epsilon]\big)
\Big] +\mathcal{O}[\epsilon^{2}]
\end{equation}
This equation is very similar to the equation obtained by authors in ref. \cite{Chaturvedi:2016kbk}
for Reissner-Nordstr\"{o}m black hole. This is not very unexpected because the charged black holes can be considered as
a special form of quintessence black holes. In the first term, $S_{BH}=\frac{r_{h}^{2}}{4G}$ corresponds to well-known Bekenstein-Hawking entropy of black hole. We can rewrite \eqref{eq:S-high-asmall} in 
terms of temperature as follows
\begin{eqnarray}
\label{eq:S-high-asmall-temp}
S^{fimite}&=&\frac{Ll}{4G}\Big(\frac{4\pi T}{3(1+\delta \omega)}\Big)^{2}(1+2\epsilon)
+\frac{L}{4G}\Big(\frac{4\pi T}{3(1+\delta \omega)}\Big)(b_{1}+\delta b_{2}) 
\nonumber \\
&+&\frac{L\epsilon}{4G}\Big(\frac{4\pi T}{3(1+\delta \omega)}\Big)
\Big[b_{3}+b_{4}\log \epsilon+\delta \big(b_{5}+b_{6} \log [\epsilon]\big)
\Big] +\mathcal{O}[\epsilon^{2}]
\end{eqnarray}
The first term scales with the area of the subsystem and represents 
thermal entropy of the region while the subsequent terms are proportional 
to the length of the boundary separating the subsystem (A) and its complement. 
 and correspond to entanglement between region $A$ and the rest of the system. 
 The quintessence corrections on the entanglement entropy is small and when $\omega \rightarrow -1$,
 these corrections approach to their maximum values. It is trivial that when $\delta \rightarrow 0$, 
 the results of Schwatschild black hole is recovered.
 \section{\label{Large charge} Black Hole in Quintessence with a Larege charege}
  In this section we explore the low and high temperature behavior of  entanglement 
 entropy for subsystem $A$ boundary to AdS black hole sourrunded by a large charge quintessence.
 The low temperature is restricted to $\omega$ colse to $-1$, therefore we first study
 the high temperature of entantnglement entropy then investigate the low temperature.
 \subsection{high temperature}
 At high temperature, the horizon radius is very large ($r_{h}l \gg 1$), 
 therefore, $r_{h}$ approach $r_{c}$. Hence, in this region the extremal 
  surface tends to wrap a part of horizon and the leading contributions come
  from the near horizon part of the surface. In other words,  when $\frac{r_{h}}{r_{c}} \rightarrow 1$, 
  the values of integrand near the horizon, $u \sim u_{0} =\frac{r_{c}}{r_{h}}$ are great and the leading
  contributions of the integral correspond to this region. Therefore, by Taylor expansion $f(u)$ around $u_{0}$, 
  the lapse function has a form as
  \begin{equation}
  \label{eq:f1-high}
  f(u) = -3(\frac{r_{h}}{r_{c}})\Big[
  1+\frac{a\omega}{r_{h}^{3(\omega +1)}}
  \Big](u-u_{0})+\mathcal{O}[(u-u_{0})^{2}]
  \end{equation}
 we can rewite this equation as
 \begin{equation}
  \label{eq:f2-high}
  f(u) \approx 3\Big[
  1+\frac{a\omega}{r_{h}^{3(\omega +1)}}
  \Big](1-\frac{r_{h}}{r_{c}}u)
  \end{equation}
  We denote the prefactor in this equation as $\delta=3[1+\frac{a\omega}{r_{h}^{3(\omega +1)}}]$. 
  This expression appears in the temperature equation \eqref{eq:Hawking-temp}. Therefore, when
  $\delta \rightarrow 0$, the temperature is low and when $\delta \rightarrow 3$, the temperature is high.
  Substituting this approximation form of the lapse function in eq. \eqref{eq:l}
, the subsystem length can be expressed as\
\begin{equation}
\label{eq:Thigh-alarge-l}
 l =\frac{2}{r_{c}\sqrt{\delta}}\int_{0}^{1}du \, 
 \dfrac{u^{2}}{\sqrt{1-u^{4}}\sqrt{1-\frac{r_{h}}{r_{c}}u}}
\end{equation}
 This equation is exactly the same derived in \cite{Chaturvedi:2016kbk} for non-extremal black hole.
  The only difference is the form of $\delta$. For computing entanglement entropy, we use  exactly the same method applied
  in \cite{Chaturvedi:2016kbk}. First by binomial expanding $\frac{1}{\sqrt{1-\frac{r_{h}}{r_{c}}u}}$,
   we obtain subsystem length as
   \begin{equation}
   \label{eq:Thigh-alarge-l2}
   lr_{c} =\frac{1}{2\sqrt{\delta}}\sum_{n=0}^{\infty}
   \dfrac{\Gamma (n+\frac{1}{2})\Gamma (\frac{n+3}{4})}
   {\Gamma (n+1)\Gamma (\frac{n+5}{4})}(\frac{r_{h}}{r_{c}})^{n}
   \end{equation}
   The series in the above equation goes as $\sim \, \frac{x^{n}}{n}$
   for large $n$, Thus it is divergent when $r_{c} \rightarrow r_{h}$.
   We isolate the divergent term. For large charge regime, 
   the horizon radius is large, therefore, we set $r_{c}=r_{h}(1+\epsilon)$.
    Substituting this expression in \eqref{eq:Thigh-alarge-l2} and expanding it
    in terms of $\epsilon$ we have
    \begin{equation}
    \label{eq:expandl-acharge}
    l r_{h}=\frac{1}{\sqrt{\delta}}\Big(
    f-\log [\epsilon] + \mathcal{O}[\epsilon]
    \Big)
    \end{equation}
       Therefore the $\epsilon$ correction can be obtained as
       \begin{equation}
       \epsilon \approx \epsilon_{ent}e^{-\sqrt{\delta} lr_{h}}
       =\epsilon_{ent}e^{\frac{-4\pi Tl}{\sqrt{\delta}}} 
       \end{equation}
    where $\epsilon =e^{f}$ anf $f$ is given by the following expression
    \begin{equation}
    \label{eq:f1}
    f=\dfrac{\sqrt{\pi}\Gamma(\frac{3}{4})}{2\Gamma(\frac{5}{4})}
    +\sum_{n=1}^{\infty}\Big(
    \dfrac{\Gamma (n+\frac{1}{2})\Gamma (\frac{n+3}{4})}
   {2\Gamma (n+1)\Gamma (\frac{n+5}{4})}-\frac{1}{n}\Big)
    \end{equation}
   Since $\frac{a\omega}{r_{h}^{3(\omega +1)}}$ is always negative,
   $\delta <3$. Thus $\epsilon$ correction is low for high temperature
   and is high for high temperature. On the other hand this correction
   increases with $a$ and $\omega$. 
   \\
    Substituting the form of the lapse function \eqref{eq:f1-high} 
    in \eqref{eq:A}, the extremal surface area can be written as
    \begin{equation}
    \label{eq:Thigh-ahigh-Aint}
    \mathcal{A}=\frac{2Lr_{c}}{\sqrt{\delta}}\int_{0}^{1}du 
    \dfrac{1}{u^{2}\sqrt{1-u^{4}}\sqrt{1-\frac{r_{h}}{r_{c}}u}}
    \end{equation}
    The integral is divergent. Regularization and computation
     are performed and finite part of extremal surface area
     is expressed as
    \begin{equation}
    \label{eq:eq:Th-al-A}
    \mathcal{A}^{finite} = Ll r_{h}^{2}(1+2\epsilon)+\frac{2Lr_{h}}{\sqrt{\delta}}
    \Big[P_{1}+P_{2}\epsilon +\epsilon \log [\epsilon]
    +\mathcal{O}[\epsilon ^{2}]\Big]
    \end{equation}
    where $P_1$ and $P_2$ are given as the following expressions
    \begin{eqnarray}
    \label{eq:P1}
    P_{1} &=&-\frac{2\sqrt{\pi}\Gamma (\frac{3}{4})}{\Gamma (\frac{1}{4})}
    +\dfrac{\log[4]-10}{8}+\frac{\pi^{2}}{6}+\sum_{n=2}^{\infty}\Big(
    \frac{1}{n-1}\dfrac{\Gamma (n+\frac{1}{2})\Gamma (\frac{n+3}{4})}
   {\Gamma (n+1)\Gamma (\frac{n+5}{4})}-\frac{2}{n^{2}}\Big)
   \\
   \label{eq:P2}
   P_{2} &=& \frac{\pi^{2}}{6}-\frac{2\sqrt{\pi}\Gamma (\frac{3}{4})}{\Gamma (\frac{1}{4})}
   -\frac{1}{4}-\sum_{n=2}^{\infty}\Big(
  \dfrac{\Gamma (n+\frac{1}{2})\Gamma (\frac{n+3}{4})}
   {\Gamma (n+1)\Gamma (\frac{n+5}{4})}-\frac{2(n-1)}{n^{2}}\Big)
   \end{eqnarray}
   Therefore the finite part of entanglement entropy will be expressed as
    \begin{equation}
    \label{eq:eq:Th-al-S}
    S^{finite} = 
    Ll S_{BH}(1+2\epsilon)+\frac{Lr_{h}}{2G\sqrt{\delta}}
    \Big[P_{1}+P_{2}\epsilon +\epsilon \log [\epsilon]
    +\mathcal{O}[\epsilon ^{2}]\Big]
    \end{equation}
    The first term scales with the area of the subsystem and is extensive. This term corresponds to
    thermal entropy of black hole. The subsequet terms are proportional to length of strip and are measure
    of the entanglement between subsystem $A$ and the rest of the system.
The effect of dark energy appears mainly in these terms as a factor $\frac{1}{\sqrt{\delta}}$.
For small $|\omega|$ and $a$, this factor is small and the effect of quintessence is also small, but 
when $|\omega|$  or $a$ goes to larger value, $\delta$ becomes small
and consequently $\frac{1}{\sqrt{\delta}}$ becomes large. when $\omega \rightarrow -1$, 
$\frac{1}{\sqrt{\delta}} \gg 1$ and the effect of dark energ will be enormous. This can be attributed to
the severe entanglement between subsystem and rest of the system at higher $|\omega|$.
   \subsection{\label{subsec:alarge-Tlow}low temperature}
   This case happens only when $\omega<-\frac{2}{3}$ especially when $\omega$ is very
   close to $-1$. For this case, 
   $\big(\frac{r_{h}}{r_{c}}\big)^{3(\omega +1)}u^{3(\omega +1)}\simeq 1$, therefore,
   the lapse function has a form as follows
   \begin{equation}
   \label{eq:alarge-Tlow}
   f(u) \approx \Big[1-\frac{a}{r_{h}^{3(\omega +1)}}\Big]
   \Big[1-\big(\frac{r_{h}}{r_{c}}\big)^{3}u^{3}
   \Big]
   \end{equation}
   This equation differs from that of pure Schwarzchild black hole in a factor
   $[1-\frac{a}{r_{h}^{3(\omega +1)}}]$ only. For $\omega =-1$ this equation converts to equation \eqref{eq:w1-A}.
   The subsystem length and minimal surface area can be obtained as
\begin{eqnarray}
\label{eq:alargew1-l}
r_{c} &\approx& \frac{2}{l}
\Big[1-\frac{a}{r_{h}^{3(\omega +1)}}\Big]^{-\frac{1}{2}}
\Big[
\sqrt{\pi}\dfrac{\Gamma (\frac{3}{4})}{\Gamma (\frac{1}{4})}
-\frac{1}{2\sqrt{\pi}}\dfrac{\Gamma (\frac{1}{4})^{3}}
{\Gamma (-\frac{1}{4})^{3}}\big(r_{h}l\big)^{3}
+\mathcal{O}[r_{h}^{6}l^{6}]\Big]
\\
\label{eq:alargew1-A}
\mathcal{A}^{finite}&\approx&\frac{L}{l}
\Big[1-\frac{a}{r_{h}^{3(\omega +1)}}\Big]^{-\frac{1}{2}}
\Big[
-\frac{\pi}{4}\dfrac{\Gamma (-\frac{1}{4})^{2}}
{\Gamma (\frac{1}{4})^{2}}
+\frac{1}{2}\dfrac{\Gamma (\frac{1}{4})^{2}}
{\Gamma (-\frac{1}{4})^{2}}(r_{h}l)^{3}
+\mathcal{O}[r_{h}^{6}l^{6}]\Big]
\end{eqnarray}   
The entanglement entropy can be expressed as
\begin{equation}
\label{eq:alargew1-S}
S^{finite}\approx\frac{L}{4Gl}
\Big[1-\frac{a}{r_{h}^{3(\omega +1)}}\Big]^{-\frac{1}{2}}
\Big[
-\frac{\pi}{4}\dfrac{\Gamma (-\frac{1}{4})^{2}}
{\Gamma (\frac{1}{4})^{2}}
+\frac{1}{2}\dfrac{\Gamma (\frac{1}{4})^{2}}
{\Gamma (-\frac{1}{4})^{2}}(r_{h}l)^{3}
+\mathcal{O}[r_{h}^{6}l^{6}]\Big]  
\end{equation}
we can see that dark energy affect the entanglement entropy as a constant factor and with increasing $a$
this effect is increased.
\section{conclusion}
In this paper we investigated analytically the entanglement entropy of a strip-like subsystem ($A$)
for the boundary field theory dual to AdS black hole surrounded by a quintessence. We obtained 
approximated analytical expression for the holographic entanglement entropy of the strip based on 
the method adopted in \cite{Fischler:2012ca}. We focused on the dependence of the 
entanglement entropy on temperature, quintessence type ($\omega$) and the amount of 
quintessence ($a$) that we called the last item as qunitessence charge. In this framework we
studied the low and high temperature of entanglement entropy for small and high $a$ regimes
and different regions of $\omega$. We find that at high temperature the entanglement entropy 
depends on temperature exponentialy and also depends on the factor $\frac{a}{r_{h}^{3(\omega+1)}}$
for small and large $a$ regimes. On the other hand, the changes of the entanglement entropy of quintessence 
black hole relative to AdS Schwarzchild black is increased when $\omega$ goes to $-1$.The behavior of entanglement entropy 
changes when $\omega$ approaches to $-1$ and this behavior has been considered for small and large $a$ regimes. 
In general the holographic tratment develops our knowledge about the entanglent and related phenomena in strongly
coupled boundary field theory at finite temperature and the charge and type of quintessence.
%%%%%%%%%%%%%%%%%%%%%%%%%%%%%%%%%%%%%%%%%%%%%%%%%%

\end{document}